___

# Software Effort Estimation using Neuro Fuzzy Inference System: Past and Present

Aditi Sharma
Department of Computer Science & Engineering
Delhi Technological University
Delhi, India
aditisharma9420@gmail.com

Ravi Ranjan
Department of Computer Science & Engineering
Dehradun Institute of Technology
Dehradun, India
raviranjan100@gmail.com

*Abstract* - Most important reason for project failure is poor effort estimation. Software development effort estimation is needed for assigning appropriate team members for development, allocating resources for software development, binding etc. Inaccurate software estimation may lead to delay in project, over-budget or cancellation of the project. But the effort estimation models are not very efficient. In this paper, we are analyzing the new approach for estimation i.e. Neuro Fuzzy Inference System (NFIS). It is a mixture model that consolidates the components of artificial neural network with fuzzy logic for giving a better estimation.

*Index Terms* - Machine Learning; Software Effort Estimation (SEE); Artificial Neural Network (ANN); Fuzzy Logic (FL); Neuro Fuzzy inference System (NFIS).

___________________________________________________*****_________________________________________________

## I. INTRODUCTION

In software Engineering (SE), development effort is represented by person-months or person-year that is, it shows the total time taken by a software development team member to perform the tasks. Software Development Effort Estimation (SDEE) is needed for strategically planning resources, team members, tasks etc. for proper allocation, for estimating the cost to develop the software, for bidding etc. It is needed for both developers and customers.

Accurate estimation of software development effort has real ramifications for the administration of software development. In the event that estimation is too low, then the product improvement group will be under impressive weight to complete the item rapidly and consequently the subsequent programming may not be completely practical or tried. Over estimation may prompt an excessive number of assets committed to extend or may prompt disappointment in contract offering.

The Standish group conducted the chaos report from year 2004 to 2012 and found that 61% of the projects worldwide were conveyed with deferral, over spending plan and numerous were not in any case wrapped up. Just 39% were delegated effective [1].

For estimating the Software Development effort (SDE), numerous procedures have been produced. They mainly fall into 3 categories:-
1) Expert judgements
2) Algorithmic Models
3) Machine Learning

Two machine learning techniques Artificial Neural Network and Fuzzy Logic, and a third which is hybrid of these two will be discussed in this paper.

## II. METHOD

The review was planned and conducted by following the Systematic Literature Review (SLR) suggested by Ketchenham and Charters [21]. The review protocol mostly incorporates six phases: research questions definition, search strategy design, study selection, quality assessment, data extraction and data synthesis.

### A. Research Questions

This survey plans to discover the Neuro-Fuzzy methods utilized for SDEE. For this, three inquiries were raised as takes after:

RQ1: Which NF strategies have been utilized for SDEE?
RQ2: What is the general estimation exactness of NFS?
RQ3: Does the traditional techniques like ANN and FL outperform the NFS?

### B. Search Strategy

This progression involves search terms, writing assets and search process.

The search terms include Software development effort estimation (SDEE), fuzzy neural network (FNN), neuro-fuzzy inference system (NFIS) and their synonyms. The search was conducted using AND, OR logic.

The writing assets for the pursuit of essential studies incorporate four electronic databases (IEEE Xplore, ACM Digital Library, Science Direct, and Google Scholar). The search terms directed already were utilized to discover the papers in these four electronic databases.

The search process comprises two phases. In the first phase, 25 papers were found from the four electronic databases, then only 19 of them were found to be relevant by following the study selection process, and in the second phase, 15 papers from the reference and of these 12 papers were added to the search list. Then study selection was done on total of these 31 papers.

*C. Study Selection*

Search phase brought about 26 papers. Since a large portion of them are not pertinent to our work so the inclusion and exclusion criteria are used for finding the relevant material, it is implemented in search phase 1 & 2 both.

Inclusion Criteria:
- Software effort estimation model of neural network and fuzzy logic for software development effort estimation.
- Review of neuro fuzzy system in SDEE using neuro fuzzy inference system.
- Use of hybrid Software effort estimation using fuzzy neural network.
- Comparative study of ANN, FL and NFS.

Exclusion Criteria:
- Estimating size, cost without estimating effort.
- Use of NFS in other fields.

*D. Quality Assessment*

In this phase, the papers are given a weightage according to their usefulness in conducting the review. In our work, we have not given them the weights, but we found by analyzing them that some of them are not very much relevant, so those papers were not considered while conducting the review. After this phase, we were left with 26 papers only for review.

*E. Data Extraction*

In this phase, the 26 papers were properly analyzed and the relevant data was extracted from them for writing the paper like comparison results etc.

*F. Data Synthesis*

The information got from the data extraction was written in this paper and the reference was given to their related data.

III. LITERATURE REVIEW

In 2011, Jioleng wen et al found in his survey that mostly eight type of ML techniques had been applied for SDEE and 26% of these studies were only on Artificial Neural Network (ANN). He likewise expressed that commonly ML strategies and Non-ML systems that were frequently used to join with other ML procedures are genetic Algorithms and Fuzzy logic respectively [2].

Recently Artificial Neural Network(ANN) and Fuzzy logic(FL) are explored to handle many problems but research on solving this problem of SDEE is very less as found by the survey conducted by Samarjeet Kaur in 2013 on 127 studies that only 10 different areas exist where NFS applications is used, but none of them was on SDEE [3]. Some of the areas in which neuro-fuzzy techniques has been implemented are pattern recognition, robotics, nonlinear system identifier and adaptive signal processing as shown by G. Bosque et al in their paper on hardware implementation and platforms in last two decades of neuro-fuzzy systems. Now we will first study ANN, FL and then NFS.

*A. Artificial Neural Network (ANN)*

ANN is a mathematical model inspired from biological Neural Network (BNN). It is a collection of processing elements which are arranged in the form of layers. A processing element corresponds to the neurons in the BNN. Based on layers, ANN is classified into 2 forms:
- Single layer Network
- Multi-Layer network containing hidden layers.

The ANN procedure begins by building up the structure of the network and setting up the strategy used to train the network utilizing a current dataset. Consequently, there are three fundamental elements: the neurons (nodes), the interconnection structure and the learning calculations [4]. ANN uses a learning process to learn from historical data and experience for providing the estimate. The most widely used learning algorithms is Back Propagation Algorithm. In this algorithm, the weighted sum of inputs is calculated and the output is generated by applying the activation function on them, then the difference between predicted and calculated value is propagated back to the network for adjusting the weights.

Neural network architecture can be isolated into two gatherings:-
1. Feed Forward networks where no loop in the network path occurs.
2. Feed backward networks that have recursive loops.

In 2013, a survey conducted by Haithen Hamza [15] shown that various ANN methods used in SDEE are:
1. Feed Forward Neural networks
2. Recurrent Neural Networks
3. Radial basis function
4. Neuro fuzzy neural network

Survey done by Jianfeng Wen [2] has calculated the MMRE (Mean Magnitude of Relative Error) to be 37% and PRED (25) (Percentage of Prediction) to be 64% while using various data sets. They likewise have called attention to that ANN can learn complex function and is additionally fit for managing noisy information. At the same time, it has weakness like
- Weak explanatory ability
- Inclined to get overfitting to the training data
- Sensitive to neural networks design and parameter setting
- Require abundant information for training

Because of these limitations, ANN is combined with other techniques by researchers for improving the estimation of accuracy. Some of these techniques are discussed in the next.

In 2010, Iman Attarzadeh [4,6] wrote a paper on ANN estimation model incorporating COCOMO2 model to overcome the uncertainties of the input parameter in the COCOMO2 model. And he validated this model using

**79**

COCOMO data set and an artificial dataset of 100 samples and found an improvement of 17.1% in MMRE in ANN-COCOMO2 model than original COCOMO2 model [4]. But when he again validated it using COCOMO and NASA93 datasets, they found only 8.36% improvement in MMRE [6]. So, even exact accuracy of an estimation model is difficult to find because different models are validated using different datasets and if the validation is done using different data sets then the same model may give different results.

In 2011, Roheet Bhatnagar presented early stage effort estimation by making SDEE at the design phase of Software Development Life Cycle (SDLC) using neural networks [7]. He validated the approach on a very small dataset based on the ER diagrams developed by engineering students as a part of their major project. The results were satisfactory, But no other researcher has validated the method on any real industry data, so not much could be said about this model.

In 2014, Mukherjee and Malu proposed a new ANN technique utilizing two layers Feed Forward Network with sigmoid hidden neurons and linear output neurons for better result and the network is trained with levenberg Marquardt back propagation [8]. Model gives better result than COCOMO, yet it needs to first choose the quantity of layers and nodes for the project. Even though, to increase the performance the quantity of layers and nodes ought to be least.

In 2015, Laqrichi, Francois, Gourc and Nevoux developed a model to associate confidence level to the Prediction Interval (PI) using a probability distribution of effort estimates [10]. In this model, first the dataset is prepared by cleaning and selecting the features and then dividing the transformed data, this data is passed to NN model structure determination for determining the appropriate NN structure based on the different design factors. And the last step i.e. Bootstrapping Neural Networks restamping is to find precise prediction interval. The researchers have used the large and most recent dataset for validating the model, and Linear Regression (LR) model shows better hit ratio than bootstrapped NN, but LR provides wider PIs, and the actual effort is found to be closer to the PI's midpoint in the new model [10]. This model should be implemented with fuzzy model, so that the PIs can give more accurate and fast results.

Most of the research over hybrid of ANN with non-ML technique was done with algorithmic models, but in 2015 Aditi, Shashank and Santanu proposed a model for enhancing the expectation precision of agile SEE (Software Effort Estimation) using different type of NNs which are

- Cascade-Correlation Neural Network,
- Group Method of Data handling(GMDH) Neural Network,
- General regression Neural Network(GRNN),
- Probabilistic Neural Network (PNN).

They validated it using twenty-one project dataset comprised of number of story points, speed of the project and the real exertion required to finish that project. While testing different NNs they found that Cascade correlation NN is best because of its self-organizing nature. But this technique could only be used if the project velocity is known. Cascade-Correlation NN has got MMRE of 14.86% and PRED of 94% [10]. So, this technique could give very clear accurate results, but as the method is only tested on a small dataset, there is no conformity of how much accurate its estimates will be if tested on large and real industrial projects.

B. *Fuzzy Logic*

Fuzzy logic (FL) model is used to provide linguistic representation for handling the uncertainties. Research on FL was first started in 1965. From that onwards, it is used in many fields as it resembles the human beings inference process. The linguistic representation is developed by allowing the processing data in form of fuzzy value having partial set relationship rather than crisp values.

The process basically comprised of 3 steps: -
1. Fuzzification- Crisp values are converted into fuzzy value by using membership function.
2. Inference Logic- A knowledge base is created, having IF-THEN type of rules and the inference engine derives the output based on these rules and the input value from first step.
3. De-Fuzzification- Here the results generated in second step is mapped to crisp values using membership function.

Various type of membership functions are available for mapping like triangular, trapezoidal MF. The precision of the estimation of Software Development effort depends vigorously on the membership functions.

There are 2 Fuzzy inference systems: -
- Namdani Fuzzy inference systems
- Takagi systems Fuzzy inference systems

Namdani expects the output to be fuzzy value whereas Takagi can have output in both fuzzy and crisp values. Because of its benefits Takagi-Sugeno Systems are implemented.

Various studies are done on these two to find the better one. One such study is given by Ram Sver P. [11] in 2015 by using 3 inputs and 1 output, when the approach is validated with 93 datasets while using triangular MF, the model has given satisfactory results having MMRE 16.85% and PRED(30) is found to be 88% [11]. Amit Sinhal and Bhupendra Verma proposed a new approach of fuzzy model using continuous Gaussian membership function and taking input as only 4 classes (which are created by categorizing the 15 features of COCOMO model) to reduce the complexity of the model. Limitation of using these techniques are that with the increase in input, the rule base also increases and for finding that rule to be fired it takes a lot of time. This model can also be tested using other membership functions as the selection of MF has a significant impact on the performance of SDEE.

80

Like Neural Network, Fuzzy logic approach is also combined with Non-ML models [12,13] by many researchers for finding more accurate estimation like Fuzzy model with COCOMO and functional point model. One such approach is given by Azzeh et al where they combine the benefits of fuzzy approach with the grey relational analysis for estimation by analogy, this model was tested for different datasets giving the PRED(25) between 48 and 67, and MMRE ranging between 23 to 52; It uses the concept of assigning different weightage to all the attributes, overcoming the need of attribute selection. FL can improve the accuracy by handling the imprecision in the input parameters to non-ML techniques like COCOMO by using some training and adoption algorithms of Software Development Effort Estimation [12].

*C. Hybrid Systems*

The advantage of using NN is its ability to learn previous data while benefit of using FL is that it processes uncertainty and represents linguistic terms making easily understandable for human brain. The advantages of both can be combined to develop a better tool. The two models can be combined in two forms:

- Fuzzy Neural Network (FNN)
- Neuro Fuzzy System (NFS)

The FNN is NN outfitted with capacity of taking care of Fuzzy data [14], While, NFS is fuzzy system having NN functionalities to improve the qualities like adaptability and versatility.

In this paper, we thoroughly studied NFS, different models who adopted NFS, benefits of NFS, limitations and comparison while other prior models conducted by other researchers. The next section will give us the answer of RQ1, i.e. which NF techniques have been used for SDEE?

IV. NEURO FUZZY INFERENCE SYSTEM

Many researchers have worked on Neuro Fuzzy Inference System i.e. the fusion of ANN and fuzzy inference system (FIS) as it has benefits of both the approaches. The fusion of two models is categorized into 3 groups [18] concurrent, cooperative and integrated Neuro fuzzy models.

*Concurrent NF Systems*: In this model, both the ANN and FIS work continuously to improve the overall performance of system. No other model helps to optimize the other model like as Cooperative NF.

*Cooperative NF Systems*: Cooperative NF systems, ANN are only used for finding the fuzzy rules from training data and then whole the work is done by FIS.

*Integrated NF System:* In Integrated NFS the learning algorithms are utilized to decide the parameters of fuzzy inference system.

The real issue with fuzzy logic is that it cannot learn and in ANN is that it is difficult to extract knowledge and it is not interpretable because of black box concept. So, to develop a model which remove both these limitations, Abraham [16] proposed two integrated Neuro- fuzzy models.

1. Mamdani Integrated Neuro- fuzzy models
2. Tokagi Integrated Neuro- fuzzy models

These frameworks are more interpretable and require less computational load but have less accuracy. Tokagi integrated NFS is a 6 layer architecture using back propagation algorithm with least mean square estimation to learn membership functions & coefficients for linear combination of the rules. The frameworks are more exact yet it likewise requires more computational exertion. As accuracy is the main concern therefore maximum newly formed techniques are based upon Tokagi integrated NFS.

The basic structure of a NFS is given by Abraham [16] having 6 layers as shown in Fig 1. To begin with layer is the Input layer in which no computation is done on the input; it is directly passed on to the next layer.

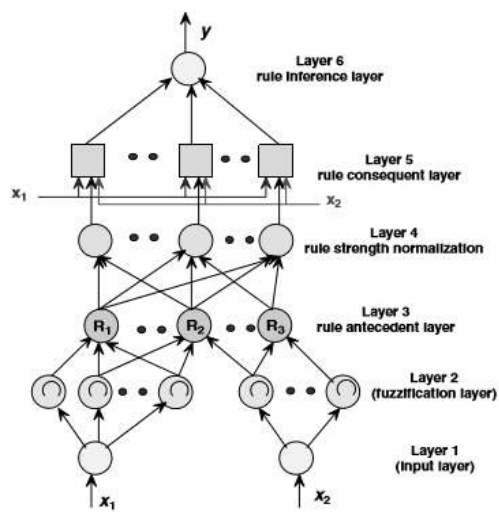

Fig. 1. Takagi Sugeno neuro-fuzzy system

Second layer is the Fuzzification layer, in which the output of first layer is converted to the fuzzy values depending on the membership function used. Third layer is the Rule antecedent layer; in this the node speaks to the precursor part of the rule applying the T-norm administrator bringing about the terminating quality of the fuzzy rule.

The fourth layer is the Rule strength normalization, in this layer the proportion of the terminating quality of the i[th] rule to the total of all the rules firing strength is calculated. Fifth layer is the Rule consequent layer, in this layer each node has a function

$$\bar{w_i} f_i = \bar{\bar{w}}(p_i x_1 + q_i x_2 + r_i) \qquad (1)$$

Where $\bar{w_i}$ is the yield of layer 4, and $\{p_i, q_i, r_i\}$ is the parameters set. A settled route is to decide the consequent parameters utilizing the minimum means squares calculation.
Sixth layer is the Rule inference layer; each node in this layer figures the general yield as the summation of every approaching sign.

81

The first integrated Neuro - Fuzzy model was ANFIS (Adaptive Network based Fuzzy Inference System) which is functionally equivalent to Takagi Sugeno type Fuzzy rule base. ANFIS was utilized for tuning a current govern base with a learning algorithm considering accumulation of training data permitting the rule base to adjust to current project [17]. ANFIS model performs well but it sometimes also produces fake rules. However, when Abraham [11] compared the ANFIS with other integrated Neuro fuzzy inference system, ANFIS was found to have the minimum root mean square error.

Other Neuro Fuzzy models have been developed in past few years; like in 2010, Wei Lin Du [26] proposed a new framework for estimating the effort having neuro-fuzzy model combined with SEER-SEM (System Evaluation of Software Resource of Software Estimation Model) to improve the accuracy of the model. In this method, 34 rating parameters and one non-rating parameter (SIBR), is given by users as the input. Rest of 34 inputs can either be a linguistic term or a continuous rating value, these are given as input to neuro-fuzzy bank which convert them to quantitative values using fuzzy sets and also providing the rules for training datasets, the output of neuro-fuzzy banks is provided as input to the SEER-SEM effort estimation which results in the effort estimation. When tested on COCOMO dataset it provides PRED (20) ranging from 15-30 depending upon whether the outliers are considered or not.

In 2012 X Huang [18] presented the framework having better interpretability because of FIS expert knowledge and traditional algorithm models. The major units of this model are Pre- Processing Neuro fuzzy system for solving the problem of interdependencies of contributing factors and to decouple them into individual factors. The second unit of Neuro fuzzy logic is Neuro-Fuzzy Bank used for adjusting the parameters of contributing factors. The third unit is the algorithm model in which any formal NL model can be used for estimating like COCOMO, SLIM, FP etc. This framework is based on divide & conquers approach. Further work on this approach is done by U.V. Sexena & S.P. Singh [19].

In 2015, Noel Garcia-Diaz et al [14] proposed a new neuro-fuzzy method having 4 MF (Gaussian Membership Function) to have better prediction and MMRE, when tested on 41 modules from 10 projects it gave an MMRE of 16.3% as compared to simple NFS having MMRE of 36%.

As compared to other techniques not much research has been done in this field. Even the different fuzzy logic models i.e. implemented using member functions or different number of input parameters are not all combined with various type of neural networks. To find the exact accuracy of these models, all should be validated using a same real project based large dataset.
*Benefits of NFS*
- Good interpretability because of fuzzy rules.
- Expert Knowledge (fuzzy logic) and learning ability of neural network can be put together.

V. COMPARISON

This section provides us the answers of RQ2 and RQ3, giving us the accuracy of NFS and showing us that it performs better than traditional techniques like ANN and FL.

The surveys conducted by Lakshmi and Binu on COCOMO dataset has found that estimate provided by Fuzzy-Neural Network(FNN) is better than the ANN and Fuzzy Logic [5].
- MMRE of ANN- 46%
- MMRE of FL- 26%-36%
- MMRE of FNN-21%-24%

And the test performed by [Venus, Amin and Luiz] on dataset of 41 projects has found that NFS is much better than ANN and FL.
- MMRE of ANN- 20.23%
- MMRE of FL- 10.57%
- MMRE of NFS-3.6%

We don't have any result on comparison between FNN and NFS till, and as these two surveys were performed on different datasets so we cannot conclude anything about these two.

CONCLUSION AND FUTURE SCOPE

Many techniques are available for estimating software development effort, but none of them is best suitable for every type of project, So, there is a need for hybrid of two or more techniques to get better accuracy. But to be certain about any new model proper testing is needed. Most of the researchers have validated the model on small and old datasets. High accuracy can be achieved if suitable model is used at proper time of software development life cycle.

Neuro-Fuzzy Inference System can be implemented with various member functions like triangle, trapezoidal, gauss etc. and with neural network implemented with various type of activation functions like sigmoid, continuous logarithmic etc. for finding best of NFIS at best time of SDLC. Neuro Fuzzy System should be tried with different algorithmic models like COCOMO, FP to test the NF model given by X. Huang, or should try to develop other hybrid models.

REFERENCES
[1] Standishgroup.com, The Standish Group ©, www.standishgroup.com,2012
[2] J. Wen ,S. Li, Z. Lin, Y. Hu, C. Huang, "Systematic Literature review of machine learning based software development effort estimation models", 2011, Elsevier
[3] S. Kar, S. Das, P.K. Gosh, "Applications of neuro fuzzy systems: A brief review and future online", 2013,Elsevier
[4] I. Attazadeh, S. H. Ow, "Proposing a new software cost estimation model based on artificial neural networks", 2010, IEEE
[5] Lakshmi R, B. Ranjan, "Survey on different machine learning techniques for software effort estimation", 2014, IEEE
[6] I. Attazadeh, A. Mehranzadh, A. Barati, "Proposing an enhanced artificial neural network prediction model to improve the accuracy in software effort estimation", 2012, IEEE

[7] R. Bhatnagar, M. K. Gosh, V. Bhattcharjee, "A novel approach to the early stage software development effort estimation using neural network models: A case study", 2011, IJCA

[8] S. Mukherjee, R. K. Malu, "Optimization of project effort estimation using neural network",2014,IEEE

[9] A. Panda, S. M. Satapathy, S. K. Rath, "Empirical validation of neural network models for agile software effort estimation based on story points", 2015, Elsevier

[10] S. Laqrichi, Francois, D. Gourc, J. Nevoux, "Integrating uncertainty in software effort estimation using bootstrap neural network", 2015,Elsevier

[11] Rama Sree P , "Analytical structure of a Fuzzy logic controller for software development effort estimation", 2015, IEEE

[12] M. A. Ahmed, M. O. Saliu, J. AlGhamdi, "Adaptive fuzzy logic based framework for software development effort estimation", 2004, Elsevier

[13] A. F. Sheta, S. Aljahdali, "Software effort estimation inspired by COCOMO and FP models : A fuzzy logic approach", 2013, IJACSA

[14] Noel G-D, Juan G-V, Nicandro F-M, Alberto V-R, Rene M-B, Evelia C-V, Leonel S-E, "Software development time estimation based on a new neuro-fuzzy approach", Information Systems and Technologies (CISTI), IEEE, 2015.

[15] H. Hamza, A. Kamel, K. Shams, "Software effort estimation using artificial neural networks : A survey of the current practices", 2013, CPS

[16] A. Abraham, "Adaption of fuzzy inference system using neural learning", 2005, Springer

[17] Jan Jantzen, "Neurofuzzy Modelling", 1998

[18] X. Huang, D. Ho, J. Ren, L. F. Capretz, "A neuro fuzzy tool for software estimation", 2004, IEEE

[19] U. V. Sexena, S.P. Singh, "Software effort estimation using neuro fuzzy approach"

[20] V. Marza, A. Seyyedi, L. F. Capretz, "Estimating development time of software project using neuro fuzzy approach" ,2008, WASNET

[21] Kitchenham and Charters, "Guidelines for performing Systematic Literature Reviews in Software Engineering", 2007, Elsevier.

[22] M. Azzeh, D. Neagu, P. Cowling, "Fuzzy grey relational analysis for software effort estimation, Journal of Empirical Software Engineering, 2009.

[23] S. Gupta, G. Sikka, H. Verma, "Recent methods for software effort estimation by analogy", ACM SIGSOFT, July 2011.

[24] G. Bosque, I. del Campo, J. Echanobe, "Fuzzy systems, neural networks and neuro-fuzzy systems: A vision on their hardware implementation and platform over two decades", Engineering applications of artificial intelligence, 2014, Elsevier.

[25] A. Sinhal, B. Verma, "A novel fuzzy based approach for effort estimation in software development", ACM SIGSOFT, September 2013.

[26] W. L. Du, D. Ho, L. F. Capretz, "Improving software effort estimation using neuro-fuzzy model with SEER-SEM ", Global journal of computer science and technology, October ,2010.